\documentclass[12pt]{article}
\usepackage{graphics}
\usepackage{epsfig} 
\usepackage{hyperref}
\usepackage{graphics}
\usepackage{color}      
\usepackage{graphicx}
\usepackage{amsmath, amssymb, graphics}

\title{\bf Thermally activated barrier crossing rate for a coupled system moving in a ratchet potential}

\author{Mesfin Asfaw$$\thanks{ Electronic address:
mesfin.taye@csun.edu}
\\
     Department of Physics and Astronomy, California State University\\ Northridge, California, USA }

\begin{document}
\maketitle

\begin{abstract}
We explore the dependence of the thermally activated barrier crossing  rate on various model parameters for a dimer that undergoes a Brownian motion on a piecewise linear bistable potential   employing   the method of  adiabatic elimination of fast variable. By introducing a different model system and approaches than the previous works \cite{c4,c5}, not only we recapture the previous results   but  we further show that systematic elimination of the fast changing variable leads to an effective Kramers type potential.  It is shown that for rigid  dimer, the escape rate $R$  monotonously   decreases with $k$. On the other hand, in the presence of time varying force, the signal to noise ratio (SNR) attains a pronounced peak at  particular  barrier height  $U_{0}$.
 \end{abstract}
 
\maketitle

 \section{Introduction}

Extended systems such as polymers and membranes exhibit challenging but fascinating transport features when they are subjected to a noisy medium and  external force. Particularly, when these polymers are exposed to a double-well potential, assisted by the  thermal background kicks, they presumably cross the potential barrier  that apparently difficult to surmount. Their flexibility and length also play a nontrivial role in the enhancement of their crossing rate as shown in the recent studies  \cite{c1,c2,c3,c4,c5,c6,c7,c8}. Furthermore,  their  escaping rate  relies  on the initial conformation of the chain. Initially stretched polymer crosses the barrier faster than a coiled chain as the coiled polymer first stretches before crossing the barrier.  The degree of stretching also  relies on the relaxation time of the polymer which itself depends on its length and coupling constant.

Numerous  studies have been also  done to explore   the response of  systems  to time varying force. In this case,  coordination of the noise with time varying force may lead to the phenomenon of stochastic resonance  (SR) \cite{c21,c22} as long as the system is exposed to a weak sinusoidal signal.  Recently  for  systems  with  more than one degree of freedom,  several studies have been conducted and showed the appearance of    SR  \cite{c9,c10,c11,c12,c13,c15,c16,c17,c18,c19,c20}.  More recently, 
we studied  the stochastic resonance  of a flexible dimer surmounting
a bistable  potential \cite{c5}. Our  numerical and analytical  analysis showed that
the SNR exhibits an optimal value   at an  optimal  elastic constant $k_{opt}$  as well as at  an optimal noise strengths $D_{opt}$.

The main objective of this paper is to explore the thermally activated barrier crossing rate  and  SR for a dimer crossing a piecewise linear bistable potential  utilizing  different model system  and approach than the previous works \cite{c4,c5}. Employing 
 the method of adiabatic elimination fast variable \cite{c5,c28,c29},   we show that systematic elimination of the fast changing variable leads to  an effective Kramers type potential. It is shown that the rate $R$   has a nonmonotonic dependence on  $k$ and $U_{0}$. Furthermore,  in the presence of time varying force,  we show that  the SNR  exhibits an optimal value at certain barrier height employing two state approach \cite{c22,c14}.  Moreover, we justify the analytic findings with numerical simulations.

The rest of the paper is organized as follows. In section II we present the model and the effective potential. In section III, we analyze the dependence  of the rate on the coupling constant and noise  strength  both analytically and  via numerical simulations.  In  section IV, employing two state approximation, we show that the SNR exhibits a maximum value at a particular $U_{0}$.  Section V deals with summary and conclusion.

\begin{figure}[h]
\begin{center}
\epsfig{file=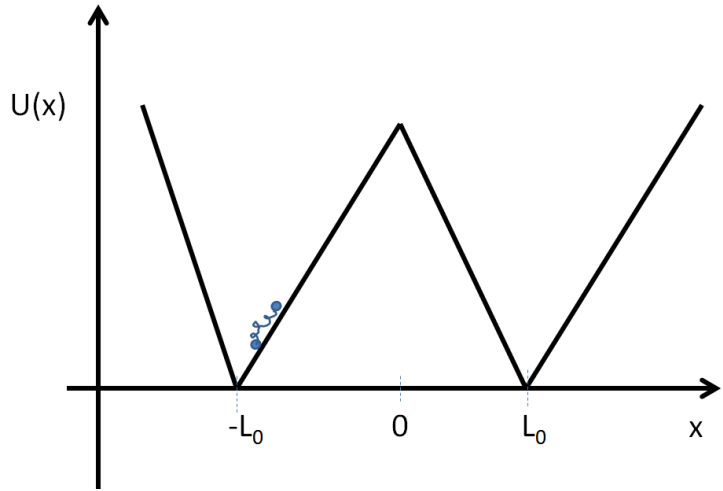,width=10cm}
\caption{Schematic diagram for a dimer  walking on a piecewise linear bistable potential. The potential wells and the barrier top are located at $x=\pm L_{0}$ and  $x=0$, respectively. The dimer is  initially situated at  $x=-L_{0}$.  } 
\end{center} 
\end{figure}

\section{The model and effective potential}
 
 We consider  a  dimer  of   harmonic chain of two beads (monomers) which undergoes a Brownian motion  along a ratchet potential where each bead has a friction coefficient $\gamma$ as shown in Fig. 1. The dynamics of the two beads is governed by the Langevin equation, 
 \begin{equation}
\gamma{dx_{1}\over dt}=-k(x_{1}-x_{2})-{\partial U(x_{1})\over \partial x_{1}}+  \xi_{1}(t)
\end{equation}
and 
\begin{equation}
\gamma{dx_{2}\over dt}=-k(x_{2}-x_{1})-{\partial U(x_{2})\over \partial x_{2}}+  \xi_{2}(t)
\end{equation}where the $k$ is the spring (elastic constant) of the dimer. 
The random force  $\xi_{n}(t)$ is considered to be Gaussian and white noise satisfying
\begin{equation}
\left\langle  \xi_{n}(t) \right\rangle =0,~~~\left\langle \xi_{n}(t)  \xi_{n}(t+\tau) \right\rangle=2D\gamma
\delta(\tau)
\end{equation}
where $D= k_{B}T$ is the strength of the thermal noise.  The rest length between the two monomer $l$ is assumed to be much less than the potential width, $l\ll 2L_{0}$ where $2L_{0}$ denotes the length of the ratchet potential.
The piecewise linear bistable potential  
energy  that the two bead experience is given by
\begin{equation}
U(x_{i})=  U_{0}({{x_{i}\over L_{0}}}+1)-2U_{0}({x_{i}\over L_{0}})\Theta(x_{i})
  \end{equation}
where $i=1,2$ and $\Theta(x_{i})$ is the Heaviside function. The potential minima  are located at  $x= \pm L_{0}$ while the    
  barrier of height $ U_{0}$   is  centered at $x=0$. For simplicity, we now  introduce dimensionless  rescaled barrier height ${\bar U_{0}}=U_{0}/D$  and rescaled length ${\bar x}=x/L_{0}$. We  also introduced a dimensionless coupling strength   ${\bar k}=kL_{0}^{2}/D$ and    time ${\bar t}=t/ \beta$ where    $\beta=\gamma L_{0}^2/D$ denotes the relaxation time. In terms of the rescaled parameters Eq. (4) is rewritten as  
$
U(x_{i})=  U_{0}({x_{i}}+1)-2U_{0}(x_{i})\Theta(x_{i}).
$

The corresponding Fokker Planck equation  for Eqs. (1) and (2) is given by 
\begin{equation}
{\partial  \over dt}P(x_{1},x_{2},t)= {\partial  \over \partial x_{1}}\chi_{1}P(x_{1},x_{2},t)+  {\partial  \over \partial x_{2}}\chi_{2}P(x_{1},x_{2},t) 
  \end{equation}
where 
\begin{equation}
\chi_{1}=k(x_{1}-x_{2})+{\partial U(x_{1})\over \partial x_{1}}+{\partial \over  \partial x_{1}}
  \end{equation}
and 
\begin{equation}
\chi_{2}=k(x_{2}-x_{1})+{\partial U(x_{2})\over \partial x_{2}}+{\partial \over  \partial x_{2}}.
  \end{equation}
In large $k$ limit, the center mass motion (CM)  $x_{cm}=(x_{1}+x_{2})/2$ of the dimer  is significantly slow compared to the relative motion $y=(x_{2}-x_{1})/2$. Our aim is here is to get rid of the fast changing variable and to write an effective  CM equation.

Before we do any analysis, next let us rewrite Eq. (5) in terms of the relative $y$ and the center of mass coordinate $x_{cm}$.  Noting that for   a rigid dimer  $U(x_{2})-U(x_{1})\approx 0$, one gets 
 \begin{eqnarray}
{\partial P\over  \partial t}= {1\over 4}{\partial \over  \partial x_{cm}}\left({\partial \over  \partial x_{cm}}\left(U(x_{cm}-y)+U(x_{cm}+y)\right)P\right)+
{1\over 2}{\partial^2 P\over  \partial x_{cm}^2} \nonumber \\
+ {1\over 4}{\partial \over  \partial y}\left({\partial \over  \partial y}\left(U(x_{cm}-y)+U(x_{cm}+y)\right)P\right) +{1\over 2}{\partial^2 P\over  \partial y^2} +\\ \nonumber {\partial \over  \partial y}\left(kyP\right).
   \end{eqnarray}
   Once again ${\partial \over  \partial y}\left(U(x_{cm}-y)+U(x_{cm}+y)\right)$ is negligible  since  for large $k$, $y= (x_{2}-x_{1})/2 \approx 0$  and hence Eq. (8)  is simplified to 
   \begin{eqnarray}
{\partial P\over  \partial t}= {1\over 4}{\partial \over  \partial x_{cm}}\left({\partial \over  \partial x_{cm}}\left(U(x_{cm}-y)+U(x_{cm}+y)\right)P\right)+
{1\over 2}{\partial^2 P\over  \partial x_{cm}^2} \nonumber \\
 +{1\over 2}{\partial^2 P\over  \partial y^2} + {\partial \over  \partial y}\left(kyP\right).
   \end{eqnarray}
 Introducing a rescaled relative term  \cite{c5}  ${\bar y}=y \sqrt{k}$ and ignoring  the bar hereafter,  the Fokker Planck equation (9) is rewritten as 
 \begin{eqnarray}
{\partial P\over  \partial t}= {\partial \over  \partial x_{cm}}\left(A'P\right)+ 
{1\over 2}{\partial^2P \over  \partial x_{cm}^2}+ 
 k{\partial \over  \partial y}\left(yP +  {1\over 2}{\partial P\over  \partial y}\right)
   \end{eqnarray}
   where $
  A'=(U'(x_{cm}-{y\over \sqrt{k}})+U'(x_{cm}+{y\over \sqrt{k}}))/4.
 $
   By integrating out the  $y$ degree of freedom, we project $P(x_{cm},y,t)$ into $P(x_{cm},t)$ as 
  $
P(x_{cm},t)=\int P(x_{cm},y,t)dy.
   $
Let us expand 
 \begin{eqnarray}
 P(x_{cm},y,t)=\sum_{n} P_{n}(x_{n},t)\psi_{n}(y,x_{cm})
   \end{eqnarray}
   where $\psi_{n}(y,x_{cm})$ is the Eigenfunction of 
   \begin{eqnarray}
  {\partial \over  \partial y} \left(y \psi_{n}(y,x_{cm}) +{1\over 2}{\partial \over  \partial y}\psi_{n}(y,x_{cm})\right)=-\lambda_{n}\psi_{n}(y,x_{cm}).
      \end{eqnarray}
 Here the eigenvalue  $\lambda_{0}=0$  and $\lambda_{n}>0$ when $n \ne 0$. Thus 
  \begin{eqnarray}
  {\partial \over  \partial y} \left(y \psi_{0}(y,x_{cm}) +{1\over 2}{\partial \over  \partial y}\psi_{0}(y,x_{cm})\right)=0.
      \end{eqnarray}
      Solving for $\psi_{0}(y,x_{cm})$,  we get 
       \begin{eqnarray}
   \psi_{0}(y,x_{cm})=\exp \left(-y^2\right) /  \sqrt{ \pi}.
      \end{eqnarray}
      
      Employing the adiabatic elimination method that explored  in the work \cite{c5}, one writes 
      the effective Fokker Planck equation as 
      \begin{eqnarray}
{\partial P_{0}(x_{cm},t)\over  \partial t}&= &{\partial \over  \partial x_{cm}}\left\{\int A' \psi_{0}dy\right\}P_{0}(x_{cm},t)+
{D\over 2}{\partial^2 \over  \partial x_{cm}^{2}}\left\{\int\psi_{0}dy\right\}P_{0}(x_{cm},t)\nonumber \\
&=&{\partial\over \partial x_{cm}}V'^{eff}(x_{cm})  P_{0}(x_{cm},t) + {D \over 2} {\partial^{2}P_{0}(x_{cm},t)\over  \partial x_{cm}^2}
   \end{eqnarray}
where  
 \begin{eqnarray}
V'^{eff}(x_{cm}) =\int_{-\infty}^{\infty}\psi_{0}(y,x_{cm})A'dy.
 \end{eqnarray}
Here   for the  ratchet potential that we considered:  
  \begin{eqnarray}
U\left(x_{cm}-{y\over \sqrt{k}}\right)=U_{0} \left(1+x_{cm}-{y\over \sqrt{k}}\right)-2 U_{0} \left(x_{cm}-{y\over \sqrt{k}} \right)     \Theta \left[x_{cm}-{y\over \sqrt{k}}\right], \nonumber
\end{eqnarray}
\begin{eqnarray}
U\left(x_{cm}+{y\over \sqrt{k}}\right)=U_{0} \left(1+x_{cm}+{y\over \sqrt{k}}\right)-2 U_{0} \left(x_{cm}+{y\over \sqrt{k}} \right)     \Theta \left[x_{cm}+{y\over \sqrt{k}}\right],  \nonumber
\end{eqnarray}
\begin{eqnarray}
U'\left(x_{cm}-{y\over \sqrt{k}}\right)=U_{0}-2 U_{0} \left(x_{cm}-{y\over \sqrt{k}}\right)  \delta\left[x_{cm}-{y\over \sqrt{k}} \right]- \nonumber 
2 U_{0}     \Theta \left[x_{cm}-{y\over \sqrt{k}}\right],  
\end{eqnarray}
and 
\begin{eqnarray}
U'\left(x_{cm}+{y\over \sqrt{k}}\right)=U_{0}-2 U_{0} \left(x_{cm}+{y\over \sqrt{k}}\right)  \delta\left[x_{cm}+{y\over \sqrt{k}} \right]-
2 U_{0}     \Theta \left[x_{cm}+{y\over \sqrt{k}}\right].
\end{eqnarray}

Substituting the above equations in Eq. (16) leads to 
  \begin{eqnarray}
V'^{eff}(x_{cm})= -{U_{0}\over 4} Erf\left[x_{cm}/{\sqrt{k}}\right].\end{eqnarray}
  
For any arbitrary small $x$, $Erf(x)\approx 2x/\sqrt{\pi}-2x^{3}/(3 \sqrt{\pi})$.  Hence $k$ is large, $V'^{eff}(x_{cm})$ is approximated as 
 \begin{eqnarray}
V'^{eff}(x_{cm})=-{U_{0}\over 4} \left(\frac{2 x_{cm}}{\sqrt{k \pi}}-\frac{2x_{cm}^3}{3 \sqrt{\pi } k^{3/2}}\right).
\end{eqnarray}

  The  effective potential  energy 
\begin{eqnarray}
V^{eff}(x_{cm})&=&\int_{0}^{x_{cm}} V'^{eff}(x'_{cm}) dx'_{cm}\nonumber \\&=& {U_{0}x_{cm}^2(-6k+x_{cm}^2)\over 24 k^{3/2}\sqrt{\pi}}
\end{eqnarray}
has potential minima  at $x_{cm}=x'_{m}=\pm \sqrt{3k}$  and maximum at $x_{cm}=0$.
The shape of the effective potential 
strictly relies on the coupling strength   and as well as on the barrier height of the ratchet potential as depicted in Figs. 2a and 2b.

\begin{figure}[ht]
\centering
{
    \includegraphics[width=6cm]{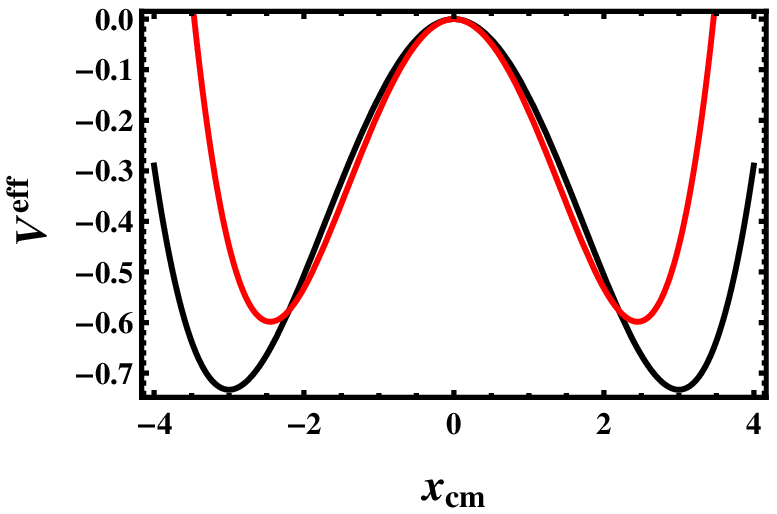}}
\hspace{1cm}
{
    \includegraphics[width=6cm]{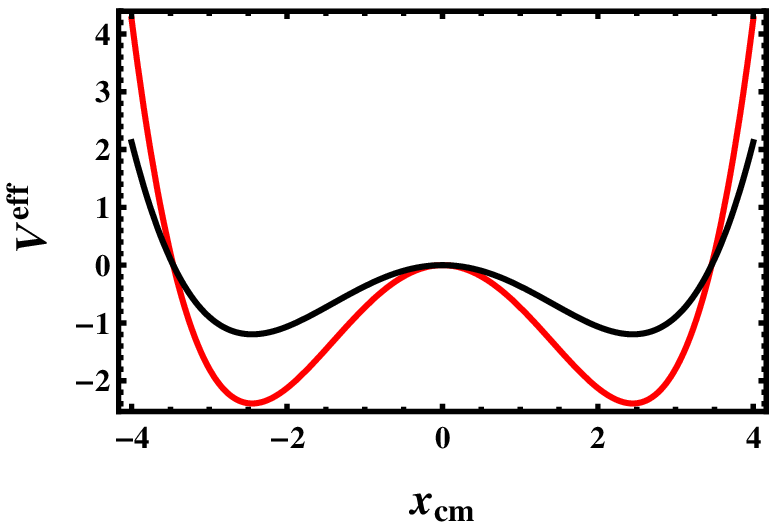}
}
\caption{(color online) (a)  The effective potential energy as a function of $x_{cm}$ for parameter choice  $U_{0}=0.5$.  The red   and black lines stand  for  $k=1$ and  $k=1.5$, respectively. (b) The effective potential energy as a function of $x_{cm}$ for parameter choice  $k=1$ (red line stands  for $U_{0}=2$ and black line for $U_{0}=1$)}. 
\label{fig:sub} 
\end{figure}

\section{The escape rate of the dimer in high barrier limit}

Consider a dimeric molecule that initially situated at the potential minima of the bistable ratchet potential.  At a frozen background temperature, the chain remains  at its initial position and   only when the temperature is above the frozen state that the dimer   undergoes unbiased random walk along the reaction coordinate. In this case, unlike a point particle, its flexibility has   nontrivial effects on its escaping  rate. The shape of the potential profile and the strength of the thermal background kicks   have also  a pronounced effect on its barrier crossing rate.  

In order to examine the various features of the  rate in the regime of large $k$, we systematically trace out the fast changing relative term which leads to an effective Kramers type potential.  For the chain that hops in the Kramers type of effective potential,   the crossing rate in high barrier limit $\nabla V^{eff}\gg
k_{B}T$ is approximated as \cite{c6} as 
\begin{equation}\label{eq:8}
        R = {\sqrt{|\omega_{0}| |\omega_{x_{m}^{'}}|}\over 2\pi} e^{-2\nabla V^{eff}}.
 \end{equation}
 where
the effective  barrier height $\nabla V^{eff}$   is the difference between  the potential energy at the saddle point and  the stable point  $\nabla V^{eff}=V^{eff}(0)-V^{eff}(-x'_{m})$. After some algebra we get 
$\nabla V^{eff}=\frac{3 \sqrt{k } U_{0}}{8\sqrt{ \pi }}$.  The parameters $\omega_{0}$ and $\omega_{x'_{m}}$ denote the curvature at the barrier top and the well minimum. We find $\omega_{0}=-U_{0}/(2 \sqrt{k\pi})$ and $\omega_{x'_{m}}=U_{0}/(\sqrt{k\pi})$. Note that limit $\nabla V^{eff}\gg
k_{B}T$ for large $k$ and $U_{0}$.

 To verify whether the results shown by the adiabatic elimination method
holds true, we compute the rate as a function of coupling
constant and barrier height  via Brownian dynamic simulations. In the simulations, an initially
coiled dimer is situated in one of the potential minima of a piecewise linear bistable potential. The trajectory for the center of mass and relative motion is simulated by considering different
time steps $t$ and time lengths $t_{max}$. The numerical accuracy is taken care
of by taking a  large number of  ensembles.

In Fig. 3a, 
we plot the rate as a function of $k$.  The dotted line is evaluated  via the numerical simulation while the solid line is  from analytical prediction.  In both cases, the rate monotonously decreases as $k$ steps up which agrees with the work \cite{c5}.  In the high barrier limit (when  $k$ is large), the simulation result approaches  to the analytic one as expected.   The dependence  of the rate on the rescaled  barrier height $U_{0}$  is   depicted  in Fig. 3b.  The simulation (dotted line) as well as the approximation (solid line) results reveal  that  the rate has an optimal value at a certain $U_{0}$.  For large $U_{0}$ (high barrier limit), both curves approach to each other.
\begin{figure}[ht]
\centering
{
    \includegraphics[width=6cm]{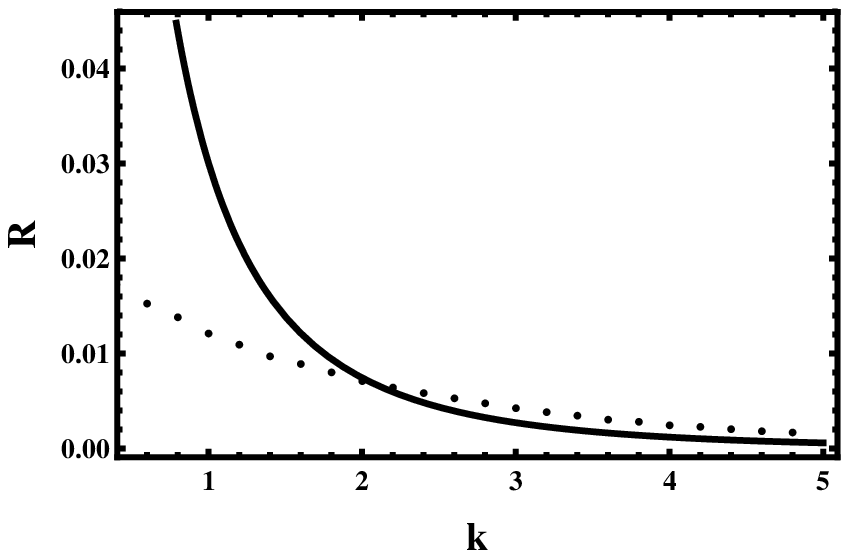}}
\hspace{1cm}
{
    \includegraphics[width=6cm]{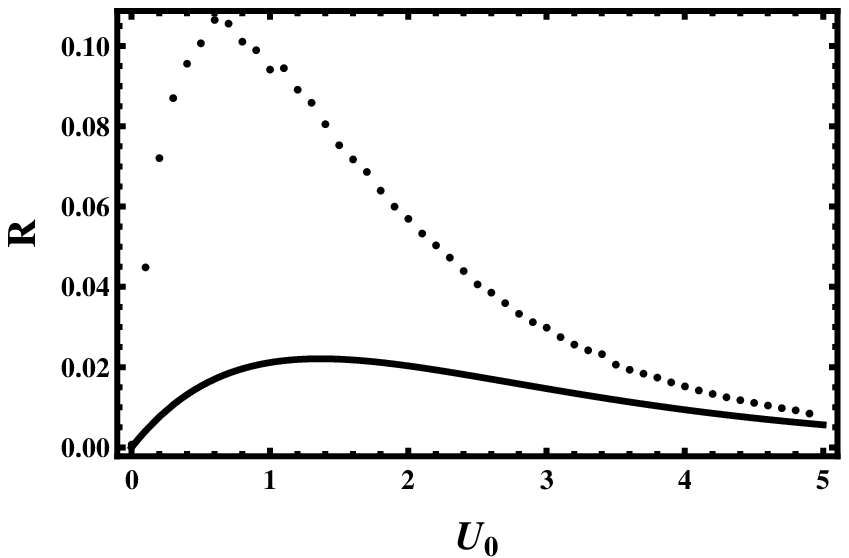}
}
\caption{(a)  Dimer crossing rate $R$  as a function of coupling constant $k$ for  fixed  $U_{0}=6.0$. The dotted line depicts the rate evaluated via   simulation  while the solid line indicates the rate from analytical prediction. (b)  The rate $R$   versus  $U_{0}$ for  parameter choice  $k=3.0$. The dotted and solid lines depict the rate  evaluated analytically and  via numerical simulation, respectively.  The rate attains an optimum value at  a certain $U_{0}$.} 
\label{fig:sub} 
\end{figure}

Note that for sufficiently small $k$, the rate $R$ exhibits an optimal value.  However, this regime cannot be  conceived by our present approach as it is only valid  in the firm $k$ regime.  At this point we want also  to stress that the flexibility of the dimer experimentally can be altered by various ways. It is known that  the hydrogen bonds firmly join the chain of the
dimer (kinesin) or polymers. Because the flexibility of the dimer is restricted
by the bond forces, breaking the hydrogen bonds may increase the rotational
degree of freedom for each atom and thereby increases the macroscopic flexibility
of the dimer \cite{a1,a2}. Moreover, hydrogen bonds, being the weakest
bonds in comparison to the covalent or ionic bonds, they can be easily broken
by thermal and chemical denaturalization which dramatically increases the
chains elasticity. The method of altering the protein flexibility via ligand
binding is also discussed in the work \cite{a4}. In addition, the increased flexibility
due to attenuated repulsions of some polymers such as DNA is reported
in the work \cite{a3}. The repulsion between phosphates along the double helix
contributes to the stiffness of the DNA. Introducing positively charged surfaces
increases the flexibility of DNA as it renormalizes the repulsion between
the two helixes. Since, the two lobes of the dimer  are mediated by flexible protein,
these all novel methods of manipulating the elasticity of the proteins
is vital for fabrication of a dimer of a specific coupling strength that can be
transported rapidly to a desired region

\section{Signal to noise ratio }

In the presence  of time varying  force,  the system reaction to the external  stimulation may depend on the flexibility of the dimer, shape of the potential profile and on the strength of the noise.
 Next  we study the dependence of the SR on these model parameters employing two state approximation for large coupling constant $k$.

In the presence of  periodic  signal  $A_{0}\cos{(\Omega t)}$, the dynamics of the system is governed by 

\begin{equation}
\gamma{dx_{1}\over dt}=-k(x_{1}-x_{2})-{\partial U(x_{1})\over \partial x_{1}}+ {\bar A}_{0} \cos (\Omega t)+ \xi_{1}(t)
\end{equation}
and 
\begin{equation}
\gamma{dx_{2}\over dt}=-k(x_{2}-x_{1})-{\partial U(x_{2})\over \partial x_{2}}+  {\bar A}_{0}\cos (\Omega t)+ \xi_{2}(t)
\end{equation}
where ${\bar A}_{0}$ and $\Omega$ denote the amplitude the angular frequency,  respectively. Here  ${\bar A}_{0}=A_{0}L_{0}/D$ and the bar will be dropped from now on.

  Employing two state model approach \cite{c22,c14}, the expression for signal to noise ratio for the  chain  has been derived in the work \cite{c22,c5}. Following the same approach, 
  for sufficiently small amplitude,  one finds
the signal noise ratio
\begin{equation} 
SNR=\pi R(2A_{0})^2
\end{equation} where
the rate $R$ is evaluated  via adiabatic elimination of fast variable. 
  
In Figure 4, the plot of SNR as a function of $U_{0}$ is presented. Both the simulation (dotted line) and the analytic (solid line) findings show that the SNR steps up with $U_{0}$ and  manifest a maximum value at a particular $U_{0}$.   Further increasing in $U_{0}$ results in a smaller SNR.  The same  figure depicts that, for large $U_{0}$, both lines approach to each other.
  
\begin{figure}[ht]
\centering
{
    \includegraphics[width=9cm]{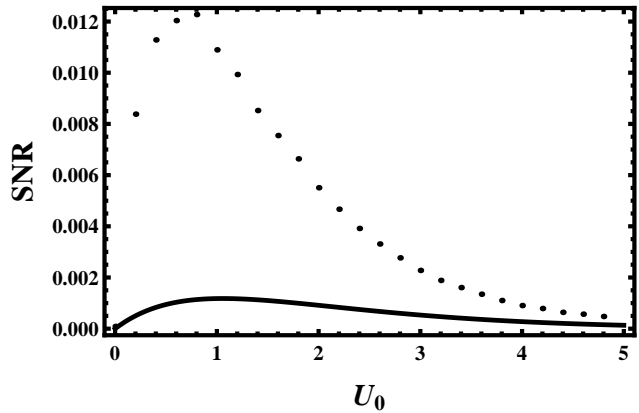}
}
\caption{ The dependence of $SNR$ as a function of noise strength $U_{0}$ for a given amplitude $A_{0}=0.1$ and $k=3$ (dotted line from simulation while solid line from analytic prediction ). The figure exhibits that the $SNR$ increases as $U_{0}$ increases and then  attains an optimal value at an optimal $U_{0}$. } 
\label{fig:sub} 
\end{figure}

   \section{Summary and conclusion}

In summary, 
 considering an initially coiled dimer which initially situated in the vicinity of  the potential minima of  a piecewise linear bistable potential, the thermally activated rate  is explored as a function a rescaled barrier height $U_{0}$.  It is shown that the rate monotonously increases with $U_{0}$ and attains a maximum value. Further increasing in $U_{0}$ leads to a smaller rate.  The  plot of $R$ as a function of $k$ also shows that  when  $k$ increases, the rate decreases. Furthermore, the response of the chain to time varying force  is  explored. In this case 
  utilizing the two state approximation, the dependence of SNR as a function  $U_{0}$  and $k$ is explored.  In this case SNR   exhibits an optimal value at optimal $U_{0}$.

This  theoretical  study is crucial since   the complicated dynamics of most of biological systems can be effectively studied by mapping into     two  coupled  bodies.  The result obtained in this theoretical work can be also checked experimentally. One
makes negatively charged dimer (coupled proteins), then put the dimer within
positively and negatively charged fluidic channel where the 
fluidic channel is subjected to an external periodic force (AC field). At low temperature, the dimer prefers to stay at a positively charged  part of the channel.  However, due to thermal fluctuations as well as  conformational change of the monomers, the dimer may presumably cross the high potential  barrier.

 In  conclusion,    since   the dynamics of complicated system such as polymer and membranes  not only relies  on shape of the  external potential  but also on their system size,  studying their dynamics is quiet complicated.  However, one can   simplify the problem  by reducing the $N$ degree of freedoms into  an effective two body problem such as dimer.  This  implies that the dimer serves as a basic model to understand the complicated dynamics of  biological systems.   Thus  we believe that   the present  theoretical study is crucial  for  the fundamental understanding of polymer and membrane physics.

\section*{ACKNOWLEDGMENTS}
We would like to thank  Yohannes Shiferaw  and Hiroshi Teramoto for discussion we had. MA would like to thank Prof. W. Sung for interesting discussions he had during his visit  at APCTP, Korea.  I would like to thank  my late mother Mulu Zebene  for the   constant encouragement.

\end{document}